# Towards early diagnosis of Alzheimer's disease: Advances in immune-related blood biomarkers and computational modeling approaches

Sophia Krix[1,2], Ella Wilczynski[3], Neus Falgàs[4], Raquel Sánchez-Valle[4], Eti Yoles[5], Uri Nevo[3,6], Kuti Baruch[5], Holger Fröhlich[1,2,*]

[1]Department of Bioinformatics, Fraunhofer Institute for Algorithms and Scientific Computing (SCAI), Schloss Birlinghoven, 53757 Sankt Augustin, Germany
[2]Bonn-Aachen International Center for Information Technology (b-it), University of Bonn, 53115 Bonn, Germany
[3]Department of Biomedical Engineering, The Iby and Aladar Fleischman Faculty of Engineering, Tel Aviv University, Tel Aviv, Israel
[4]Alzheimer's Disease and Other Cognitive Disorders Unit, Neurology Service, Hospital Clínic de Barcelona, FRCB- Institut d'Investigacio Biomedica August Pi i Sunyer (IDIBAPS), University of Barcelona, Barcelona, Spain
[5]ImmunoBrain Checkpoint Ltd., 3 Haim Pekeris St., Rehovot 7670203, Israel.
[6]Sagol School of Neuroscience, Tel Aviv University, Tel Aviv, Israel

\* Correspondence:
Holger Fröhlich
holger.froehlich@scai.fraunhofer.de

**Keywords:** Alzheimer's Disease, biomarkers, early diagnosis, immune system, modeling, agent-based modeling, machine learning, omics

**Abstract**

Alzheimer's disease has an increasing prevalence in the population world-wide, yet current diagnostic methods based on recommended biomarkers are only available in specialized clinics. Due to these circumstances, Alzheimer's disease is usually diagnosed late, which contrasts with the currently available treatment options that are only effective for patients at an early stage. Blood-based biomarkers could fill in the gap of easily accessible and low-cost methods for early diagnosis of the disease. In particular, immune-based blood-biomarkers might be a promising option, given the recently discovered cross-talk of immune cells of the central nervous system with those in the peripheral immune system. With the help of machine learning algorithms and mechanistic modeling approaches, such as agent-based modeling, an in-depth analysis of the simulation of cell dynamics is possible as well as of high-dimensional omics resources indicative of pathway signaling changes. Here, we give a background on advances in research on brain-immune system cross-talk in Alzheimer's disease and review recent machine learning and mechanistic modeling approaches which leverage modern omics technologies for blood-based immune system-related biomarker discovery.

# 1 Introduction

Alzheimer's disease (AD) is prone to have a strongly increasing prevalence worldwide due to the aging society. The number of individuals living with AD is estimated to triple from 2019 to 2050, rising from 57.4 million to 152.8 million (GBD 2019 Dementia Forecasting Collaborators, 2022). Deaths due to AD have increased in recent years, making AD one of the top leading causes of death in many countries around the globe ("2023 Alzheimer's Disease Facts and Figures," 2023; Barton & Saib, 2023; Ren et al., 2022). The onset of the disease likely starts decades before the first symptoms of cognitive decline appear, when changes in amyloid-beta or tau in the cerebrospinal fluid (CSF) as well as in positron emission tomography (PET) measurements become detectable (Mattsson-Carlgren et al., 2020). According to the Amyloid/Tau/Neurodegeneration (ATN) framework (Jack et al., 2018), levels of $A\beta_{42}/A\beta_{40}$ ratio, P-tau181 and neurofilament light in the CSF determine the diagnosis of AD biologically. PET neuroimaging and/or CSF-based measurements are both costly or invasive procedures that are only performed at specialized centers with long waiting lists. On the other hand, screening cognitive tests available at primary care are frequently not sensitive enough to identify the disease in the stage of mild cognitive impairment (MCI). In conclusion, there are currently limited resources for an early diagnosis of AD.

This situation contrasts with the requirements of recently available treatment options, in particular aducanumab and lecanumab (Budd Haeberlein et al., 2022; van Dyck et al., 2023), which received regulatory approvals by the US Food and Drugs Administration (FDA) in 2021 and 2023, respectively. Their effectiveness has only been demonstrated in individuals with early AD, diagnosed with mild MCI or mild dementia due to AD. Given that these treatments are only beneficial in patients in an early stage of AD, it is unclear how a larger fraction of patients could benefit from the novel therapeutic options within the current healthcare setting. It is thus essential to come up with easy to use and reliable diagnostic procedures that could help to identify subjects at these early clinical stages.

One option in this regard are blood-based surrogates of amyloid and (phosphorylated) tau which have been developed in recent years. For example, in a recent study levels of $A\beta_{42}/A\beta_{40}$ ratio, P-tau181 and neurofilament light in the blood demonstrated comparable prediction performance for conversion to AD in the next 4 years to the state-of-the-art CSF biomarkers, yielding the best results when omitting the amyloid-beta variable (Cullen et al., 2021). In 2020, the first blood test for AD by C2N Diagnostics got approval by the FDA (*C2N Earns CE Mark for Alzheimer's Blood Test*, 2020), which measures Aβ42/40 levels in the blood and the ApoE proteotype. In combination with age, the test yields a score that was evaluated on its potential to determine the status of amyloid plaques in the brain (Hu et al., 2022). This was followed by an amyloid plasma panel by Roche which measures pTau 181 and APOE E4 in 2022 (*Roche Gets Breakthrough Device Status for Alzheimer's Test*, 2022). Recently, Roche has announced to collaborate with Eli Lilly for the enhancement of the plasma panel for the early detection of AD (*Roche Announces Collaboration with Lilly to Enhance Early Diagnosis of Alzheimer's Disease*, 2023). However, these blood-based tests for the diagnosis of AD only consider a combination of amyloid beta and phospho-tau as proxies for the levels of the same proteins in CSF.

As a reaction to the recent developments in plasma biomarkers, the Alzheimer's Association have revised the criteria for diagnosing and staging AD and included plasma biomarkers in their proposal (Alzheimer's Association, 2023). In particular, biomarkers of inflammatory and immune processes are incorporated as a new category in the proposed framework, which reflect reactivity of astrocytes and microglia. The recognition of emergent plasma biomarkers as valuable extensions of the pre-existing ATN framework is a first step forwards to a more accessible and earlier diagnosis of AD.

In recent years it has become increasingly clear that there is a cross-talk between the immune cells of the central nervous system and the periphery (Kipnis et al., 2004; Ziv et al., 2006; Rolls et al., 2007; Baruch & Schwartz, 2013; Kunis et al., 2013; Baruch et al., 2013), which is also reflected in the number of research findings on AD and immune system-related genes (see Suppl. Figure 1). This opens new perspectives to search for immune based blood biomarkers.

The purpose of this article is to a) shed light on the recent findings on the role of the immune system in Alzheimer's disease, and b) to point out the connection with computational approaches, including machine learning and prospectively also mechanistic modeling, for the successful identification of immune based blood biomarkers.

## 2  Brain-Immune System Crosstalk in Alzheimer's Disease

Recent anatomical and functional discoveries of ways by which the immune system interacts with the brain have shaped the long-held dogma of the brain as an immune privilege organ. Instead of being considered as solely secluded behind the blood-brain-barrier from having any interaction with the peripheral immune system, it is becoming increasingly clear that the central nervous system (CNS) is in constant dialogue with immune cells, and that these interactions are taking place at unique anatomical barriers, such as the choroid plexus (Baruch et al., 2014; Baruch & Schwartz, 2013) and the meningeal spaces (Louveau et al., 2018; Rustenhoven & Kipnis, 2022).

This life-long crosstalk shapes brain function in health and disease, and was repeatedly implicated in AD development and progression (Bettcher et al., 2021). For example, various studies in transgenic mouse models of AD have shown that increasing recruitment of myeloid cells to the brain is associated with reduced amyloid pathology and improved cognitive performance (Baruch et al., 2015; Ben-Yehuda et al., 2021; El Khoury et al., 2007; Koronyo et al., 2015; Naert & Rivest, 2013; Rosenzweig et al., 2019). While the mechanism of action is not fully understood, it seems that bone-marrow derived myeloid cells that enter the brain perform distinct roles, compared to the local microglia, in reducing neuroinflammation and toxic protein pathology.

### 2.1  Immune Cellular Biomarkers

Microglia, the resident immune cells of the brain, play a central role in maintaining brain homeostasis. They are the first line of defense against invading pathogens and also respond to injury or disease states by becoming activated and performing tasks such as clearance of cellular debris and dead cells. Microglia have been shown to play a dual role in AD. In the early stages, they can limit Aβ accumulation by phagocytosing these peptides and promoting their clearance. This neuroprotective role of microglia is substantiated by numerous studies demonstrating that stimulation of microglial activity can result in reduced Aβ burden and improved cognitive function (Miao et al., 2023). However, prolonged activation of microglia can lead to a chronic inflammatory state, causing neurotoxicity and contributing to neuronal death (Block & Hong, 2005). Indeed, microglial activation is a common feature in AD brains, and is often associated with increased levels of pro-inflammatory cytokines (W.-Y. Wang et al., 2015).

The role of infiltrating monocytes, another key component of the myeloid cell lineage, has also been explored in the context of AD. Recent evidence suggests that these cells can be recruited to the brain in response to Aβ deposition, where they differentiate into macrophages and contribute to Aβ clearance (Shechter & Schwartz, 2013). Interestingly, monocytes appear to be more efficient at clearing amyloid-beta (Aβ) than resident microglia (Fani Maleki & Rivest, 2019), which may be due to their phagocytic capabilities. Indeed, cell surface marker analysis of myeloid cells that enter the brain, in comparison to microglia, show distinct expression of scavenger receptors, such as MSR1 (Frenkel et al., 2013), and ability to phagocytose Aβ plaques (Malm et al., 2008; Simard et al., 2006) or reduce soluble Aβ oligomer pathology (Naert & Rivest, 2012). Another intriguing aspect of myeloid cell function in AD is their potential role in tau pathology. Tau is a microtubule-associated protein that becomes hyperphosphorylated and forms neurofibrillary tangles in AD. Recent studies in tauopathy mouse models have shown that recruitment of blood monocytes to the brain is associated with a reduction in tau pathology (Ben-Yehuda et al., 2021; Rosenzweig et al., 2019). However, the precise mechanism by which these cells influence tau pathology is still not well understood, and further studies are needed to elucidate their role. Particularly, it is not clear if the beneficial effect is directly mediated by the recruited myeloid cells, or an indirect effect of the reduction in neuroinflammation.

T cells, a key component of the adaptive immune system, represent another layer of complexity in the brain-immune crosstalk in AD. The brain was thought to be devoid of T cells under non-pathological conditions, but recent studies have challenged this notion. It has been discovered that T cells are present in the meningeal spaces and choroid plexus under normal conditions and were suggested to play a role in brain function (Evans et al., 2019; Filiano et al., 2017; Schwartz & Baruch, 2014). Immunophenotyping analysis of blood samples from AD patients have shown significant reduction in T cell frequencies (Richartz-Salzburger et al., 2007), and a recent meta-analysis of 36 studies showed that this reduction is associated with increased CD4/CD8 T cell ratio in patients with AD compared to HCs (Huang et al., 2022). Beyond these general changes in lymphocytes, consistent findings have shown changes in numbers and phenotype of various T cell subsets in AD (Dai & Shen, 2021).

One intriguing line of research is investigating the role of T cells in AD pathology. T cell presence was repeatedly demonstrated in post-mortem human brain tissue of persons with AD (Merlini et al., 2018; Togo et al., 2002), and in the cerebrospinal fluid (Lueg et al., 2015; Monson et al., 2014). Outside the brain, peripheral T cells in the blood show reduced frequencies in AD patients (Huang et al., 2022). In mice, brain infiltration of T cells was studied in the context of their spatial distribution in the CNS-borders, such as the meninges, and brain parenchyma, and was shown to correlate with the degree of Tau pathology and to contribute to neurotoxicity (X. Chen et al., 2023). However, the role of T cells in promoting or suppressing AD pathology is unclear, and most likely involve different subsets of T cells. Indeed, T cell deficiency in AD mice was shown to inhibit hippocampal neurogenesis and restrict hippocampal neuronal regeneration (J. Liu et al., 2014).

Of potential interest is immune checkpoint molecules as cellular biomarkers in AD. Programmed cell death protein 1 (PD-1) and its ligand PD-L1 are key immune checkpoint molecules that play a crucial role in regulating immune responses. They are typically expressed on the surface of T cells and other immune cells and serve to dampen immune responses, preventing autoimmunity and maintaining self-tolerance. Expression of PD-1 on T cells and PD-L1 on monocytes and macrophages significantly decreases in AD patients and in patients with MCI compared with age- and sex-matched healthy controls (Saresella et al., 2012). In a recent study, this change in PD-1/PD-L1 expression on T cells was correlated with the different stages of AD (Wu et al., 2022). PD-1 expression was also found to increase on T cells in the cerebrospinal fluid of AD patients (Gate et al., 2020). Beyond the potential use of PD-1/PD-L1 as cellular biomarkers in AD, this immune checkpoint pathway was also suggested as a target for therapeutic intervention. In different mouse models of AD, transient blockade of PD-1/PD-L1 resulted in reduced brain pathology and improved cognitive performance (Baruch et al., 2016; Rosenzweig et al., 2019). The mechanism of action was shown to involve homing of specialized immune cells to the brain (both myeloid cells and regulatory T cells), where these cells mitigate different pathomechanisms, ultimately leading to function improvement (Ben-Yehuda et al., 2021; Dvir-Szternfeld et al., 2022, p.).

## 2.2 Response in Microglia and Astrocytes

Neuroinflammation is a physiological consequence of injury of the CNS due to a given insult. This mechanism intends to provide neuroprotection and repair neuronal damage in post-injured tissues. Thus, a cascade of events derived from neuronal damage triggers inflammatory responses in the CNS through the activations of glial cells and the release of cytokines and growth factors. Glial cells are abundant in the brain featuring different characteristics, functions, and phylogenetic origin. Some of them, such as the astrocytes (neural origin) and microglia (differentiated blood monocytes), play prominent roles in homeostasis and neuroinflammatory processes (Valles et al., 2023).

Nevertheless, despite the initial positive effect of the neuroinflammation on the post-injured tissue, this mechanism can eventually become detrimental to neuronal homeostasis and associated processes (Singh, 2022). Chronic or imbalanced inflammatory responses driven by aging may contribute to and perpetuate the physiopathology of neurodegenerative diseases, including AD. Although the specific immune cross-talk between microglia, astrocytes, and neurons is still a matter of discussion, many studies suggested that the neuroprotective mechanisms of glia/astrocytes turn neurotoxic by interacting with beta-amyloid promoting senile plaques and tau accumulation through a cascade of pro-inflammatory mediators, such as the release of nitric oxide and cytokines, which eventually contribute to neuronal death. Due to this cascade of astrocytic and microglial activation, immune-derived molecules are released, allowing its measurement as specific soluble markers for astrocytic and microglial activity.

### 2.2.1 Astrocytes

Astrocytes are the most present glial cell of the CNS, derived from neural stem cells maintaining homeostasis and providing metabolite and growth factors to neurons. Besides, astrocyte is central in synapse formation and plasticity, modulating the extracellular balance of ions and removing free radicals. Under pathological conditions, astrocytes undergo morphological and functional changes leading to cell hypertrophy (reactive astrocytes) and increased release of neurotoxic factors. In AD, reactive astrocytes aggregate in the vicinity of Aβ plaques, as post-mortem and rodent studies show, suggesting a direct interaction between Aβ and astrocytes (Wegiel et al., 2001). Besides, astrocytic activation modifies protein expression, such as the glial fibrillary acidic protein (GFAP) and chitinase-3-like protein 1 (YKL-40), both soluble astrocytic makers typically measured in patients with AD.

#### 2.2.1.1 Glial fibrillary acidic protein (GFAP)

Several studies have demonstrated a significant increase in GFAP levels measured by CSF and plasma in AD compared to control individuals even in preclinical stages of the disease (Ishiki et al., 2016; Oeckl et al., 2019). Importantly, clinical studies support that plasma measures of GFAP correlate with its CSF levels, suggesting that plasma measures are a robust proxy of astrocyte reactivity in the brains of living individuals (Benedet et al., 2021). In addition, it has been shown that GFAP levels in both CSF and plasma correlate with cognitive and biological measures of AD progression. For instance, elevated plasma and GFAP levels are associated with lower cognitive performance and steeper cognitive decline (Fukuyama et al., 2001; Saunders et al., 2023). Moreover, recent studies have shown that higher GFAP plasma levels correlate particularly with amyloid burden measured both by CSF and amyloid-PET quantification in opposition to Tau burden (Pereira *et al.*, 2021; Ferrari-Souza *et al.*, 2022; Bucci et al., 2023). Furthermore, Aβ pathology has been associated with increased plasma phosphorylated tau levels only in individuals positive for astrocyte reactivity (i.e., elevated GFAP), suggesting a modulating role of astrocytic activity between Aβ and tau pathology in AD (Pascoal et al., 2023).

#### 2.2.1.2 Chitinase-3-like protein 1 (YKL-40)

A meta-analysis including 14 cohorts demonstrated a significant increase in YKL-40 CSF levels in AD compared to cognitively unimpaired controls (Bellaver et al., 2021). Moreover, YKL-40 levels differ between neurodegenerative dementias (Antonell et al., 2020). Along the same line, some studies measuring plasma YKL-40 showed increased AD participants with a good correlation with CSF levels (Craig-Schapiro et al., 2010). Besides, a recent study showed that YKL-40 was positively associated with memory performance and negatively associated with brain

Aβ deposition, suggesting a potentially protective effect of glia on incipient brain Aβ accumulation and neuronal homeostasis (Vergallo et al., 2020).

### 2.2.2 Microglia

Microglia are the brain resident macrophages derived from monocyte precursor cells significantly involved in immune defense and homeostasis maintenance. In AD, Aβ aggregates induce microglia morphological and molecular changes in microglia overexpression of particular receptors, cytokines, and other factors. Clinical and preclinical studies have shown that activated microglia have a prominent role in AD progression. Nevertheless, there is a discrepancy on whether microglia-mediated inflammation could be neurotoxic or neuroprotective, which may depend on the timing, duration, and amplitude of microglia activation. The interest for measuring microglia activity using soluble markers has gained interest in the field of AD, being the Triggering receptor expressed on myeloid cells 2 (TREM2) most studied.

#### 2.2.2.1 Triggering receptor expressed on myeloid cells 2 (TREM2)

TREM2 mediates the interactions between Aβ and microglia. TREM2 is a receptor expressed on the surface of microglia that regulates Aβ degradation and clearance by binding to Aβ and bringing it to the microglia's lysosome (Heckmann et al., 2019). The expression of the TREM2 gene in AD could change microglia's response to Aβ, as a loss of functional TREM2 in microglia leads to a decreased microglial clustering and increased Aβ seeding, suggesting its control over Aβ pathology. Recent evidence in individuals with and without AD pathology showed that increased levels of CSF TREM2 were associated with the slower amyloid accumulation and tau (measured by PET scans) and cognitive decline, highlighting the protective functions of microglial in AD (Pereira et al., 2021). CSF sTREM2 levels are highly correlated with plasma sTREM2 (A. Zhao et al., 2022). Peripheral (plasma) sTREM2 is altered in AD in a stage-specific manner, being more altered in AD than in MCI individuals (Weber et al., 2022; A. Zhao et al., 2022). Moreover, lower TREM2 levels are associated with altered peripheral immune response in AD, including other inflammatory factors such as fibroblast growth factor-2, GM-CSF, or IL-1β (Weber et al., 2022).

#### 2.2.2.2 Galectin-3

Galectin-3 (Gal-3) is a beta-galactosidase binding protein involved in microglial activation. In opposition to TREM2, Gal-3 seems to have a deleterious role in AD. It is primarily expressed around Aβ plaques in both human and mouse brains, and knocking out Gal-3 reduces AD pathology in AD-model mice (García-Revilla et al., 2022). Compared to controls, CSF Gal-3 levels are elevated in AD patients and correlate with tau and synaptic markers (GAP-43 and neurogranin) instead of amyloid-β (Boza-Serrano et al., 2022). In addition, it is associated with other CSF neuroinflammatory markers, including sTREM-2, GFAP, and YKL-40 (Boza-Serrano et al., 2022). Studies including CSF and serum measurements of Gal-3 in AD or other neurodegenerative diseases showed similar results (Ashraf & Baeesa, 2018). Moreover, Gal-3 levels are progressively increased across the AD stage and are associated with reduced global cognitive outcomes (MMSE) (X. Wang et al., 2015, 2019).

Together they suggest that astrocyte and microglia biomarkers are consistently altered in AD and likely reflect an essential role in AD physiopathology. Given their potential as biomarkers tracking neuroinflammatory processes in AD and their principle accessibility through blood testing, further investigation is needed. Improving its understanding would contribute to generating consensus on using astrocyte biomarkers in AD and including them in the AD clinical research framework.

## 3     Omics-Based Biomarker Signature Discovery and Machine Learning

Omics-based approaches should allow for an unbiased, data-driven discovery of immune-based blood-biomarkers in Alzheimer's Disease. However, classical bulk RNA sequencing is limited due to the fact that mixtures of various cell types are measured. Going one step further, recent single cell sequencing techniques now allow the measurement of the expression of all genes in individual cells - including immune cells in the blood. This provides a more detailed picture of a patient's disease state at a given point in time and may help to obtain a better understanding of disease mechanisms as well as associated biomarkers in the future. For example, Xu and Jia analyzed single cell gene expression data of 3 AD and 3 controls, and their findings suggest that the peripheral adaptive immune response, mediated by T cells, is a factor in the pathogenesis of AD (Xu & Jia, 2021). In a study by Xiong et al. on scRNA sequencing of AD patients, they identified B cells as a determinant of the severity of the disease. A lower number of B cells was found in the blood of AD vs healthy controls, and in a follow-up experiment in mice they found a correlation between depletion of B cells in early stage AD models and accelerated cognitive decline as well as increased amyloid-beta burden (Xiong et al., 2021). Similar results were gathered in a study by Song et al., where additionally an increase in proportion of neutrophils as well as gene expression levels of AD-associated pathways in neutrophils were detected via a cellular deconvolution method on RNA bulk blood sequencing data (Song et al., 2022).

The richness in information of single-cell RNA sequencing data does not come without challenges. High dimensionality, technical noise and batch effects impose difficulties for data processing and require the use of specific computational tools for further analysis, which have been extensively developed in recent years (Garmire et al., 2021; Hao et al., 2021; Zappia et al., 2018).

Classically, omics data are analyzed using statistical analysis methods, which helps to provide insights into disease mechanisms and candidate biomarkers. However, statistical methods only help to understand differences between patient groups that exist on average and do not allow us to make statements about a single patient. But to support medical decisions, including early diagnosis, we need to consider and combine features of an individual patient into a diagnostic score. Since it is unlikely that a single biomarker would allow for a highly accurate identification of patients in a preclinical or prodromal stage, for this purpose machine learning (ML) plays a crucial role (Fröhlich et al., 2018). The application of ML bears huge potential for biomarker discovery since it enables us to find patterns in high dimensional data that are otherwise hard to detect. ML might help us to understand the heterogeneity of the disease and aid in identifying subtype-specific biomarker signatures. In particular, precision medicine takes individual patient characteristics on a molecular level into account and can help to identify biomarker panels that allow for a precise diagnosis. In the following we review existing works focusing on the discovery of immune-based blood biomarkers signatures using machine learning. For an overview of the studies, we refer to Table 1, listing approaches based on their primary data source type with information on the data set and the machine learning method. Detailed information on the immune system-related blood biomarkers identified in each study that were found to distinguish between AD, MCI and HC individuals can be found in Supplementary Table 1.

### 3.1 Blood-based Proteomics- and Transcriptomics Approaches

In an extensive multi-center study, Morgan et al. investigated plasma biomarkers for the diagnosis and stratification of AD patients (Morgan et al., 2019). More than 50 inflammatory proteins were measured in immunoassays, including complement components, activation products and regulators, cytokines and chemokines. A logistic regression model was trained to differentiate between the diagnosis groups enabled to identify inflammatory biomarkers that distinguished not only HC from AD, but also MCI from AD, with an AUC of 0.79 and 0.74, respectively. Plasma analytes in AD patients compared to HC were increased for C4 and eotaxin-1, and decreased for CR1, C5 and CRP. In MCI patients, increased levels of FH, C3 and MCP-1, and decreased levels of C5 and MIP-1-Beta compared to HC were found. When comparing AD to MCI, increased eotaxin-1 and MIP-1-Beta, and decreased FI, C3, CRP and MCP-1 emerged as distinguishing plasma biomarkers.

Prabhakar and Bhargavi investigated in a machine learning approach to identify which blood plasma proteins could be useful in the early detection of AD (Prabhakar & Bhargavi, 2022). The blood plasma protein samples were analyzed for a total of 146 protein features. The best biomarker panel was identified by feature selection and subsequent combination testing with Support Vector Machines. The panel consists of the following immune-related proteins: A2M, MDC, IL-18 and CD5L. This set of biomarkers achieved an accuracy ranging from 0.55 to 0.8, depending on the kernel used, for the classification of AD and HC. When applied to MCI individuals, this protein panel did not yield high accuracy results.

Karaglani et al. developed diagnostic biosignatures based on transcriptomics and proteomics from 7 public datasets for the diagnosis of AD (Karaglani et al., 2020). With proteomic data consisting of nearly 1000 features, they identified 7 protein biomarkers using logistic regression with an AUC of 0.921, including several immune-system related proteins, such as CAMLG, IL-4, TPM1 and IL-20.

Jammeh et al. investigated the best combination of blood biomarkers for a routine diagnosis of AD-related dementia from blood samples from the ADNI proteomic database (Jammeh et al., 2016). They identified a panel of biomarkers, including immune-related proteins A1M, A2M, eotaxin-3, PYY, PPY and EGF via a Naïve Bayes classifier, which was able to identify AD patients with a sensitivity of 0.85 and a specificity of 0.78. In a further study, Eke et al. tried to define an optimum panel of blood biomarkers that can fulfill a diagnostic performance of 80% sensitivity and specificity (Eke et al., 2018). They used proteomic data from the ADNI phase 1 study. With Support

Vector Machines, they could identify 5 biomarkers associated with the immune response, comprising A1M, A2M, C3, and TNC, that could sufficiently distinguish AD from HC in the ADNI cohort.

Choi et al. investigated the association of protein levels of CypA, HO-1 and IRE1 in the blood with changes in gray matter volume (Choi et al., 2021). They found that in both MCI and AD individuals, blood levels of all three proteins investigated were correlated with AD signature regions of the brain. Higher CypA levels were associated with increased gray matter volume of the occipital gyrus and posterior cingulate. Gray matter volume changes at the hippocampus, uncus, lateral globus pallidus and putamen were positively associated with changes in HO-1 levels and negatively with IRE1 levels (Choi et al., 2021).

In a study by Liu et al., several subsets of immune cells were quantified in AD individuals (Z. Liu et al., 2021). They found a significant increase in immune infiltrates in AD individuals compared to HC, such as monocytes, M0 macrophages, and dendritic cells, and a decrease in other immune cell types, such as NK cell resting, T-cell CD4 naive, T-cell CD4 memory activation, and eosinophils. They identified hub genes, which include ABCA2, CREBFR, CD72, CETN2, KCNG1 and NDUFA2 by applying LASSO regression and SVMs to bulk RNA gene expression profiles. A further analysis of the identified hub genes and their relation to immune factors through the TISIDB database confirmed that these genes were strongly correlated to the level of immune cell infiltration and regulators of the immune microenvironment (Ru et al., 2019).

Walker et al. developed a protein signature for dementia risk based on the dysregulation of immune and autophagy pathways in middle-aged adults (K. A. Walker et al., 2023). They analyzed 4877 plasma proteins of 10.901 individuals from the ARIC cohort in terms of their association with dementia risk up to 25 years later. They found 32 dementia-associated proteins, of which 12 were related to CSF biomarkers of AD, neurodegeneration or -inflammation. They grouped the identified plasma proteins into modules based on protein coexpression patterns and found associations of several modules with near-term or long-term dementia risk. Modules associated with long-term dementia risk were enriched for proteins involved in JAK-STAT signaling, T helper 1 and 2 cell differentiation, leukocyte activation and immune/mitogen-activated protein kinase signaling.

A study by Abdullah et al. aimed to find the transcriptomics biomarkers that could most accurately classify AD patients in Malaysia (Abdullah et al., 2022). They used Boruta's feature selection algorithm on a transcriptomics dataset from the TUA study, comprising 22.254 transcript genes of 92 AD patients and 92 HC. They evaluated the classification performance on several statistical and machine learning classifiers. With an elastic net logistic regression model, they achieved an accuracy of 0.82. Among the 16 potential biomarkers that they identified were ANKRD28, CCDC92, DEFA3, FBXO32, GRIA4, HDAC7, IFITM3, LY6G6D, MC1R, RPL18, SPOCD1, ST14, TOR1AIP2, TRIM16L, UBXN7, and VEGFB.

Kim and Lee proposed a pathway information-based neural network for the prediction of AD, which uses blood and brain transcriptomic signatures (Kim & Lee, 2023). They used pathway information from KEGG and Gene Ontology alongside gene expression data as input for their deep neural network. With the help of a backpropagation-based model interpretation method, they were able to identify essential pathways and genes in the prediction of AD. This analysis indicated an enrichment of genes involved in PI3K-Akt and MAPK signaling, two pathways which are involved in the immune response (Hawkins & Stephens, 2015; Y. Liu et al., 2007), in association with AD.

The link between neuroinflammation and AD has gained more attention in recent years. Since biological findings attribute both a detrimental as well as protective role to neuroinflammation (Weitz & Town, 2012), Gironi et al. proposed to use an approach that can reconstruct non-linear relationships to model the complex system of neuroinflammation. In their study, Gironi et al. analyzed the immunological and oxidative stress parameters in peripheral blood mononuclear of AD and Mild Cognitive Impairment patients (Gironi et al., 2015). They constructed a machine learning algorithm to distinguish healthy controls from AD and MCI patients and selected the most important immunological and oxidative stress parameters for the prediction. They concluded that the initial activation of microglia is beneficial for amyloid clearance but that this mechanism can become chronically destructive when it is not timely controlled.

The cross-talk between the immune system and mitochondrial dysfunction has been studied recently, where a mitochondrial-related gene expression signature has been identified that is in interaction with the immune microenvironment (Zhang et al., 2023). The identified differentially expressed hub genes are correlated with several immune cell types, including memory B cells, effector memory CD8 T cells and natural killer T cells.

With the help of text-mining, functional enrichment analyses and protein-protein functional interaction analyses, it was possible to explore the molecular network associated with neuroinflammation in AD computationally (El Idrissi et al., 2021). In their study, El Idrissi et al. identified key regulatory proteins of neuroinflammatory processes in AD. Their findings are relevant for research in diagnostics of AD and potential treatments.

### 3.2 Blood-based Autoantibodies

The role of autoimmunity in neurodegenerative diseases, including AD, has seen more attention recently, and offers new perspectives in terms of diagnostics and therapeutics (Kocurova et al., 2022; Lim et al., 2021; Meier-Stephenson et al., 2022; Prüss, 2021). Autoantibodies are antibodies that react to self-antigens, and are all-present in the human body. Natural autoantibodies are responsible for clearance of debris during inflammation, yet they might also amplify inflammation in systemic auto-immune and neurodegenerative diseases (Elkon & Casali, 2008; Sardi et al., 2011; Shim et al., 2022). Therefore, several approaches have taken up this idea and investigated blood-based autoantibodies as biomarkers for the diagnosis of AD.

DeMarshall et al. tried to identify biomarkers for patients diagnosed with MCI due to an early-stage AD pathology using autoantibodies (C. A. DeMarshall et al., 2016). They selected 50 MCI patients and their HC from the ADNI2 study and performed protein microarrays on the serum samples to identify autoantibodies. They used a Random Forest model to identify a panel of 50 AD-associated MCI-specific biomarkers. They reported that their model could differentiate MCI patients from age- and gender-matched controls with a sensitivity, specificity and accuracy of 100%, and furthermore, with > 90% from mild-moderate AD. In a further study, DeMarshall et al. used this panel of autoantibody biomarkers on elderly hip fracture repair patients (C. DeMarshall et al., 2019). With their autoantibody panel, they were able to identify the patients that were positive for CSF AD biomarkers. Recently, DeMarshall et al. proposed a multi-disease diagnostic platform based on autoantibodies to detect the presence of AD-related pathology, focusing on early stages, including the pre-symptomatic, prodromal and mild-moderate stages (C. A. DeMarshall et al., 2023). They conclude that blood-based autoantibodies present an accurate, non-invasive, low-cost solution, especially for early diagnosis of AD in pre-symptomatic and prodromal AD stages.

## 4   Mechanistic Modeling of the Immune System

The immune system is dynamic and characterized by complex cell-cell interactions, which eventually manifest in the increase or decrease of certain markers over time. Therefore, purely data driven statistical and machine learning approaches, which typically rely on cross-sectional snapshot data, often lack robustness and reproducibility across studies (Forouzandeh et al., 2022). A principal alternative is thus to first come up with a detailed, quantitative understanding of fundamental disease mechanisms, from which biomarker candidates may then be derived and tested in a second step (Generalov et al., 2019; Hampel et al., 2019; Zewde, 2019). Here, mechanistic modeling techniques could fill in a gap by simulating longitudinal data on cell-cell interactions based on parameter estimates from quantitative data. Mechanistic modeling approaches, and specifically agent-based modeling (ABM) techniques, have been developed in the past to provide a realistic simulation of mixtures of various cellular species, accurately describing cytokine concentrations, activations of cells and interactions between cellular players of the immune system over time. In the following we provide an overview about existing works focusing on the modeling of immune related mechanisms in the AD field.

### 4.1   Ordinary Differential Equations (ODEs)

Ordinary differential equations (ODEs) are widely used to mathematically describe time-dependent molecular processes in systems biology (C. Chen et al., 2014; Elowitz & Leibler, 2000; Heinrich et al., 2002; Jain et al., 2011; Khodayari & Maranas, 2016; Mannan et al., 2015; Morken et al., 2014; Neuert et al., 2013; Wilkinson, 2009). Often experimental data obtained from cultured cells is used to define the initial conditions of such models and to infer free parameters. The ready fitted model can then be used for extrapolation of time series or for simulating counterfactual scenarios, for example the intervention by a certain drug. Learning the structure of such a mechanistic model can give biological insights and may also point towards candidate biomarkers. Recently, several mathematical models on AD progression have been published that take the role of the immune system into account, which we will elucidate in the following.

A kinetic model was used to explore the effect of microglia and astroglia on the pathogenesis of AD (Thuraisingham, 2017, 2018). The model suggests that these immune cells promote the progression of AD via neuronal cell death. They argue that the increase in population of microglia and astroglia in AD means an increase of inflammatory cells producing toxins that eventually cause neuronal cell death. Along these lines, the aggregation of microglia in AD was proposed to be explained via a chemotaxis model (H.-Y. Jin & Wang, 2018). They modeled the interaction of several attractive and repulsive cytokines produced by microglia in order to explain under which conditions the aggregation of microglia is possible, as seen in the pathology of Alzheimer's patients. With their model, they were able to explain the aggregation of microglia given a certain combination of chemotactic responses of microglia to IL-1beta and TNF-alpha.

Several convergent mechanisms exist in Alzheimer's and Parkinson's disease, including the p38 pathway activation that enhances the production of proinflammatory cytokines such as IL-1beta and TNF-alpha (Corrêa & Eales, 2012). Sasidharakurup et al. used a systems biology tool to create process diagrams of common mechanisms, and converted each reaction into a mathematical equation dependent on an initial literature-derived condition. They conclude that the activation of microglia is responsible for increased levels of TNF-alpha, as observed in Alzheimer's and Parkinson's disease, which lead to excessive oxidative stress and result in necrosis and apoptosis (Sasidharakurup et al., 2020).

Kyrtsos and Baras developed a graphical systems biology model to investigate the interaction of neuroinflammation, mitochondrial function, the ApoE genotype and A-β generation on cellular and molecular level (Kyrtsos & Baras, 2013). They simulated chronic low-level inflammation by increasing the level of TNF-alpha and simulated the effects on the levels of various neuroinflammatory cytokines. Interestingly, together with a triggered collapse of mitochondrial function, chronic neuroinflammation led to neuronal cell death. Their model results agree with biological findings which have shown a decreased capability of the brain to protect itself in case of chronic inflammation (Newcombe et al., 2018).

During the progression of AD, a strong accumulation of CD4+ T cells is seen in many patients (Monsonego et al., 2013). CD4+ T cells can regulate immune responses via secretion of signaling molecules, yet the set of cytokines produced by each CD4+ T cell can vary depending on the cytokines in the extracellular environment. The duration of T cell receptor engagement and co-stimulation also contributes to the differentiation of CD4+ T cells. Miskov-Zivanov et al. simulated the changes in cell fate and plasticity of CD4+ T cells with a logical circuit model (Miskov-Zivanov et al., 2013). In their review on heterogeneity and function of CD4+ T cells, Carbo et al. gather computational approaches to model immune responses of CD4+ T cells and we refer to them for further reading (Carbo et al., 2014). The relationship between antigenic stimulations of CD4+ T cells and regulatory CD4+ T cells was recently described in an ODE model (Yusuf et al., 2020), using their concentration and the extent of the antigenic stimulation. ODE models typically neglect spatial aspects of biophysical mechanisms. Sego et al. therefore combined non-spatial ODE modeling with spatial, cell-based modeling via a cellularization approach to create a spatiotemporal model of the immune response upon viral infection (Aponte-Serrano et al., 2021; Sego et al., 2022).

Notably, most ODE approaches use initialization parameters and relative rates from the literature, hence they do not reflect individual system-level differences. These approaches can rather be seen as a generalization of immunological processes, but are not suitable for patient-specific inductive reasoning.

### 4.2 Agent-based Modeling and Cell-Cell Interaction Models

Cell-cell interactions are influential on organismal development and single-cell function (Armingol et al., 2021). The understanding of cell-cell interactions can give insight into biological mechanisms in development of disease. Agent-based modeling (ABM) is a powerful technique to simulate and explore phenomena that include a large number of active components, represented by agents. In the ABM framework the agents are operating in the system, simultaneously influencing the simulated environment and being influenced by the simulated environment. The agents can also perform actions autonomously, based on rules or state machines, with regards to their interaction with other agents and with the environment (Efroni et al., 2003). These actions represent the behaviors in the real system (Adra et al., 2010; Kaul & Ventikos, 2015; Sun et al., 2009; Vodovotz et al., 2004). Modeling of a multi-scale spatiotemporal system is a complex computational challenge due to the high number of sub-processes involved, each with its own features. As the complexity of the simulated system increases in size and in the agents' capabilities, the outcome of the simulation may reveal unpredicted emergent results that were otherwise very hard to obtain, e.g. by pure mathematical modeling (Bauer et al., 2009; Narang et al., 2012). These emergent results can be later interpreted as specific signaling pathways at the intra-cellular level thus inferring on biomarkers discovery (D. C. Walker & Southgate, 2009). ABM requires mainly local knowledge regarding the mechanisms (rules) that govern the behavior of each type of agent and the environment, whereas global behavior emerges from the agent-agent interactions as well as the interactions with the environment.

Lately, there have been extensive research efforts to develop agent-based simulation systems. General purpose approaches include EPISIM (Sütterlin et al., 2009, 2013), SimuLife (Bloch et al., 2015), CellSys (Hoehme & Drasdo, 2010), and several approaches have been developed to specifically model the immune system, such as the Multiscale Systems Immunology project (Mitha et al., 2008), LINDSAY Composer (Sarpe & Jacob, 2013), FLAME (Tamrakar et al., 2017, 2017), Simmune (Meier-Schellersheim et al., 2006), C-ImmSim (Baldazzi et al., 2006) and Cell Studio (Liberman et al., 2018a, 2019). Agent-based models tailored towards specific immune system processes also exist, such as a recently published model of adaptive immunity that describes the T cell response to various factors, including antigen-presenting dendritic cells, changes in T cell recruitment and swelling conditions of the lymph node (Johnson et al., 2021).

Cell Studio (Liberman et al., 2018b, 2019) is a unique platform for modeling complex biological systems. It provides an advanced environment specifically designed for non-coding researchers, including a visual interface, modeling of biological, biophysical, bioinformatics and chemical data, as well as parallel computing. In particular, the platform is specialized in modeling immunological response at the cellular level. Cell Studio's main feature is to realistically model the immunological response as a multi-scale, hierarchical phenomenon that operates at molecular, cellular, tissue levels and eventually the organ level. The platform, adopting the ABM paradigm, proposes the cell as a native agent whose interactions with proteins, molecules, medium and other cells, define the main features of the immunological scenario. The choice of which interactions are necessary to describe a particular process is given to the user which can set up a unique experiment with predefined rules. To facilitate the definition process, a state-machine description of the cell is used, to control the dynamic behavior as an approximation for the intracellular processes, in line with the growing number of publicly available pathways and network databases.

Next to agent-based modeling techniques, several cell-cell communication and interaction inference approaches exist. These rely heavily on prior knowledge from pathway databases and use various methods to estimate cell-cell interactions which can be categorized into statistics-based, network-based and spatial transcriptomics-based approaches (Z. Liu et al., 2022). Obviously cell-cell interaction models and ABM are not necessarily separated. However, grossly speaking, cell-cell interaction models relate to networks of interactions and attempt to be very comprehensive in integrating omics data to include a multitude of possible interactions. The majority of ABM approaches have a similar aim, but they include additional dimensions of the interactions as not covered by most cell-cell interaction models such as space (including cellular motion, diffusion and mechanical characteristics of the medium), temporal kinetics of secretion, specific effects of different concentrations and finally the ability to include variance within same species of cells. Inclusion of this physical data comes with the price of a need to simplify the modeled interactions due to the complexity of parameterization and, later-on, of computation.

Dimitrov et al. compared current cell-cell communication prediction approaches (Dimitrov et al., 2022). Their analysis showed that immune system pathways are not equally represented in the resources used for the approaches under investigation. They found that both the interaction information resource as well as the method can considerably impact the cell-cell communication inference prediction, and conclude that integration of information from additional modalities could help to refine the predictions.

Several cell-cell communication approaches have been specifically designed to integrate single-cell RNA sequencing data (Armingol et al., 2021; Shao et al., 2020), which open up a new way to generate patient-specific models for cell-cell communication prediction. Since there is no ground truth for the cell-cell interaction prediction, Liu et al. suggest to use spatial transcriptomics data for the evaluation of cell-cell interaction models with single-cell RNA sequencing data and provide a comparison of approaches (Z. Liu et al., 2022).

One of the first to integrate patient-level data of Alzheimer patients into cell-cell interaction simulations was Zhao et al., using the first published single-cell RNA sequencing dataset of peripheral blood on Alzheimer's with 3 affected individuals (Xu & Jia, 2021). They analyzed cell-to-cell communication networks using CellChat (S. Jin et al., 2021) to gain insight on signaling pathways and interactions between different cell subpopulations in Alzheimer patients as well as ovarian cancer patients (S. Zhao et al., 2023). They found that interactions between monocytes, natural killer cells and T cells are increased in AD individuals compared to healthy controls, and that monocytes were highly influenced by HLA-related signals. They followingly designed a monocyte-based prognostic signature which was used to determine the risk of a patient, and validated their risk model on ovarian cancer datasets. They successfully predicted low- and high-risk group progression and survival via Cox analysis with the risk score. Similar experiments now need to be performed on AD datasets to validate the suitability of such a monocyte-related risk model for the prediction of progression and survival of individuals affected by AD. This is one prestigious example of how cell-cell interaction models can be useful in the development of a prognostic biomarker signature based on immune signaling and cell-cell interactions.

Along these lines, simulating the cross-talk between the peripheral immune system and the brain for the condition of Alzheimer's disease via agent-based models might open up new ways to find immune-specific biomarkers that are indicative of early changes in the immune system during the onset of the disease. Monitoring changes in biomarker concentrations and expression levels during the course of the ABM simulations might give insights into which biomarkers might be more or less suitable to identify a patient at an early stage. Going one step further, the integration of patient-specific information as part of the initialization of the ABMs could help to shed light on the differences in biomarker levels on an individual level and thereby enable personalized predictions on the intensity and velocity of changes in biomarkers. This information could be helpful in the identification of disease mechanisms specific to the phenotype of the patient and aid in better design of patient-specific treatments.

## 5 Conclusion

Identifying patients in a preclinical or prodromal disease stage of AD is vital for the success of currently available therapeutic options. Hence, alternative and easily accessible biomarkers which allow for an early diagnosis are urgently needed. Current knowledge suggests that the ATN framework for diagnosis of AD should be complemented with markers of inflammation. Given the recently discovered cross-talk of immune cells of the central nervous system with those in the peripheral immune system, immune-based blood-biomarkers seem to be a promising option. However, it is unlikely that a single biomarker alone would allow for an early diagnosis with sufficiently high accuracy. The focus should be on the discovery of biomarker signatures, for example using modern omics technology, which allow sequencing immune cells in the blood. Machine learning algorithms can then combine and integrate signals from various genes into a single, patient-specific diagnostic score, which could enable identifying patients at an early disease stage.

Data-driven biomarker discovery approaches are limited by the often observed lack of statistical robustness and reproducibility due to limited sample size and - in case of omics data - extremely high dimensional feature space. Mechanistic modeling approaches, including ODEs and ABMs, emerge as complementary methods in this regard. These techniques allow for high statistical robustness, but as they are knowledge-derived, the unknown or not completely understood aspects of the disease can be misrepresented. Moreover, most mechanistic models have so far only been calibrated and tested against experimental data in model systems (e.g. cell lines), but never been applied to real patient data. Recent ABMs, such as Cell-Studio, could close this gap of purely knowledge-derived features, as they are initialized with patient-level data, for example FACS measures of cell surface markers, which reflect the observed mixture of different cell types in an individual patient at a given time, thus allow for predictive results. In the future ABMs could potentially also be initialized by single cell RNA seq data, hence providing a simulation of the dynamical behavior of gene expression. ABM platforms require an elaborate work of physical and chemical parametrization that typically demands very significant human resources. It is likely that the next generation of parameterization in ABM will be facilitated by automated processing of literature by natural language processing and generative transformers. Analyzing the simulated dynamics from ABM in combination with modern data mining and machine learning techniques might open new perspectives for biomarker signature discovery. In particular, the consideration of a time dimension may open the possibility to identify biomarker signatures in a far more robust and statistically stable manner than currently (Diaz-Uriarte et al., 2022).

# 6   Conflict of Interest

KB and EY received salaries from ImmunoBrain Checkpoint (IBC). RSV reports consultancy or speaker fees from Ionis, AviadoBio, NovoNordisk, Pfizer, Neuraxpharm, Roche diagnosis and Lilly.

# 7   Funding

This project is supported in Germany through the BMBF (Funding reference 01ED2206A), in Spain through the ISCIII (Grant number AC21_2-00007 to RSV) co-funded by the EU, under the aegis of the EU Joint Programme – Neurodegenerative Disease Research (JPND) – www.jpnd.eu